\documentclass[a4paper,10pt,twocolumn,twoside,journal]{article}

\usepackage{geometry}
\advance\oddsidemargin by -1in
\advance\evensidemargin by -1in
\advance\headsep by 2.8mm
\usepackage{listings}
\usepackage{braket}
\usepackage{pgfplots}
\usepackage{subcaption}
\usepackage{bm}
\usepackage{placeins}

\usepackage{cite}
\usepackage{hyperref}
\usepackage{url}
\usepackage{authblk}
\usepackage{mathptmx}
\usepackage{booktabs}
\usepackage{amsthm}
\usepackage{amssymb,amsmath,mdwlist,mathtools}
\usepackage{multirow}

\def\BibTeX{{\rm B\kern-.05em{\sc i\kern-.025em b}\kern-.08em
    T\kern-.1667em\lower.7ex\hbox{E}\kern-.125emX}}

\usepackage[normalem]{ulem}

\providecommand{\keywords}[1]{\textbf{\textit{Index terms---}} #1}

\title{Multi-Class Quantum Convolutional Neural Networks}

\author[]{Marco~Mordacci}
\author[]{Davide~Ferrari}
\author[]{Michele~Amoretti}
\affil[]{\small Quantum Software Laboratory, University of Parma, Parma, Italy}

\date{}

\begin{document}

\maketitle

\thispagestyle{empty}
\pagestyle{empty}

\begin{abstract}
  Classification is particularly relevant to Information Retrieval, as it is used in various subtasks of the search pipeline.
  In this work, we propose a quantum convolutional neural network (QCNN) for multi-class classification of classical data. The model is implemented using PennyLane. The optimization process is conducted by minimizing the cross-entropy loss through parameterized quantum circuit optimization. The QCNN is tested on the MNIST dataset with 4, 6, 8 and 10 classes. The results show that with 4 classes, the performance is slightly lower compared to the classical CNN, while with a higher number of classes, the QCNN outperforms the classical neural network.
\end{abstract}



\section{Introduction}
Research on machine learning for Information Retrieval (IR) has grown rapidly in the last few years~\cite{Onal2017,Zamani2022,Salemi2023}. The advent of quantum computers contributed to a further expansion of the field, leveraging Quantum Machine Learning (QML) algorithms~\cite{Cerezo2022,Mensa2023,Guarasci2022}. 
Specifically, Quantum Neural Networks (QNNs)~\cite{tacchino2019artificial} have shown to be able to achieve a significantly better effective dimension than comparable classical neural networks~\cite{Abbas2020,Coles2021}.

Particular attention has been given to classification tasks, using hybrid quantum-classical approaches. 
Within IR, classification is used in various subtasks of the search pipeline: preprocessing, content filtering, sorting, ranking, and more.
Since the seminal work by Tacchino et al.~\cite{tacchino2019artificial} that introduced a quantum binary-valued perceptron, many other attempts have been made to demonstrate quantum advantage in binary or multi-class classification tasks~\cite{Chalamuri2021,Zhang2022,Silver2022,Zhou2023}.

In this work, a quantum convolutional neural network (QCNN) that is able to perform multi-class classification is presented. 
A general theoretical framework is proposed, where classical data are encoded into quantum states and processed by means of quantum computations, and the parameters of the quantum model are updated according to measurement results. The classifier, which is based on the general QCNN architecture introduced by Cong et al. in ~\cite{cong2019quantum}, consists of convolutional layers composed of two-qubit gates and pooling layers to reduce the number of qubits within the network. The parameterized quantum circuit is trained using classical machine learning methods by minimizing the cross-entropy loss through gradient descent optimization of the circuit parameters. A preprocessing circuit is introduced to improve the performance of the model.

A performance evaluation of the proposed QCNN, considering the MNIST dataset for the classification of handwritten digits, is reported. It is shown that, with 6, 8 and 10 classes, the proposed QCNN is more accurate than a classical CNN of comparable complexity. Furthermore, the QCNN achieves results comparable to those in \cite{bokhan2022multiclass} but with significantly fewer parameters. Additionally, it effectively tackles classification problems involving 6 and 8 classes, delivering satisfactory outcomes. 

The paper provides the following new contributions:
\begin{itemize}
    \item the use of a preprocessing filter before the convolutional filters, to enhance expressibility and entanglement;
    \item simulation results showing that the QCNN outperforms, in terms of accuracy, the classical CNN when executed with 6, 8, and 10 classes, when training is based on the full MNIST dataset;
    \item simulation results showing that the proposed QCNN still outperforms the classical CNN, in terms of accuracy, when training is based on a significantly reduced number of samples of the MNIST dataset.
\end{itemize}

The paper is organized as follows. In Section~\ref{sec:related}, some related works are discussed. In Section~\ref{sec:framework}, the theoretical framework is described. In Section~\ref{sec:results}, the application of the proposed framework to the MNIST dataset is presented, with simulation results that compare the quantum model to a classical model of equivalent complexity. Finally, Section~\ref{sec:conclusion} concludes the paper with a discussion of future work.

\section{Related Work}
\label{sec:related}

In the last few years, several quantum neural networks for binary classification have been proposed. On the other hand, for the multi-class scenario, there are few proposals. 

In their seminal work, Tacchino et al.~\cite{tacchino2019artificial} introduced a quantum algorithm implementing a quantum binary-valued perceptron, showing exponential advantage in storage resources over alternative realizations. The model can achieve an advantage over classical perceptron models in the classification of 4 and 16 bit strings. Furthermore, Mangini et al.~\cite{Mangini2020} extended the perceptron model to classify greyscale images. It is tested on the MNIST dataset taking into account only images of zeros and ones, for which it achieves an accuracy of $98\%$.

In~\cite{Chalamuri2021}, Chalumuri et al. designed a quantum multi-class classifier with a variational quantum circuit consisting of one layer of data encoding followed by three layers for each time step: $R_{y}$ gates on data qubits, $CNOT$ gates to entangle data qubits with each other and with ancillary qubits, $R_{z}$ and $R_{y}$ gates on ancillary qubits. The data encoding is performed through a layer of $H$ gates on the data qubits and a unitary $U(x_{i})$ for feature $x_{i}$ on qubit $q_i$ such that $U(x_{i}) \frac{\ket{0}+\ket{1}}{\sqrt{2}}= \cos{x_{i}}\ket{0}+\sin{x_{i}}\ket{1}$.
The authors show that with their encoding scheme, they can achieve better accuracy with respect to amplitude encoding, up  to $50\%$ increase on test datasets with a $91\%$ accuracy, on standard datasets with three classes.

In~\cite{Silver2022}, Silver et al. present Quilt, a framework for multi-class classification designed to work with current noisy quantum computers. Quilt uses Principal Component Analysis (PCA) as preprocessing to fit data on small quantum computers and encodes such preprocessed data onto a quantum state with the well known amplitude encoding algorithm. The core feature of Quilt is the employment of not just one classification circuit, but multiple variants of small classification circuits, called ensembles. The goal is to achieve better accuracy by weighting and aggregating the results of each small circuit. When the ensemble has a low confidence on the classification result, a OneVsAll classifier is deployed as a support network to reach a final decision. The authors show that Quilt can achieve an accuracy of almost $80\%$ with the MNIST dataset and eight classes.

In~\cite{hur2022quantum}, Hur et al. introduced a QCNN-based classifier for binary classification of MNIST digits and MNIST Fashion dataset. They devised a QCNN that solely relies on two-qubit interactions throughout the entire algorithm, following a similar approach to~\cite{cong2019quantum}. The results achieved about $99\%$ of accuracy on MNIST dataset and $94\%$ on MNIST Fashion dataset. The proposed network does not provide accurate results when used with more than 2 classes.

In~\cite{bokhan2022multiclass}, Bokham et al. present a QCNN-based classifier for 4-class classification. The circuit consists of three components: the quantum encoding, which is implemented using amplitude encoding, the convolutional layer and the pooling layer. The authors show that their network can achieve accuracy between $85\%$ and $90\%$ on the MNIST dataset, depending on the selected classes and an accuracy between $85\%$ and $93\%$ on the fashion MNIST dataset. However, the QCNN does not outperforms the classical CNN in terms of accuracy.

In~\cite{wei2022quantum}, Wei et al. present a QCNN framework consisting of an initial state preparation layer, through amplitude encoding, a convolutional layer, comprising multiple multi-controlled parameterized operators and a pooling layer. Finally, after the pooling, a fully connected layer measures the remaining qubits. The authors show that while the implemented QCNN can reach almost $80\%$ accuracy with 4 classes of the MNIST dataset with a gate complexity speedup with respect to a CNN, it never surpasses the CNN.

\section{Theoretical Framework}
\label{sec:framework}
The goal of an L-class classification problem is to assign the correct label to an unseen data point $x \in \mathbb{C}^N$, given a labelled data set
\begin{equation}
    D = \{(x_1, y_1), (x_2, y_2), ..., (x_M, y_M)\}.
\end{equation}

To solve the multiclass classification problem, a parameterized quantum circuit $U(x_i, \theta)$ is constructed, which is generally denoted as Quantum Neural Network (QNN). The QNN is trained by optimizing the parameters of the quantum gates in order to minimize a cost function $C(\theta)$.
After the training is completed, a test set is used to assess the network's generalization capability.

The QNN is composed by the following parts: the encoding circuit, which encodes the classical data into quantum states; the variational quantum circuit, whose parameters $\theta$ need to be trained; a measurement, that allows the network to predict a class given an input image, and an optimization process implemented classically, which has to update the parameters $\theta$ of the variational quantum circuit for minimizing the cost function $C(\theta)$. 

\begin{figure*}
    \centering
    \includegraphics[width=1.0\textwidth]{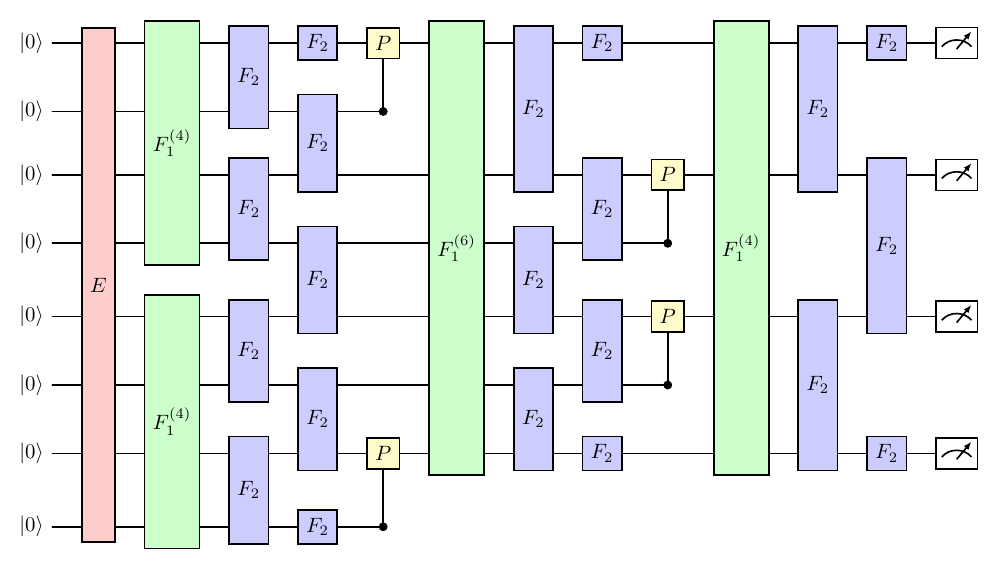}
    \caption{General structure of the quantum convolutional neural network for 10-class classification. It consists of several steps: encoding of the classical data ($E$), preconvolutional filters ($F^{(n)}_1$), convolution filters ($F_2$), pooling layer ($P$) and measurement.}
    \label{Fig:qcnn}
\end{figure*}

In the following, the parts of the proposed QNN are described in detail. It will be clear that the designed variational quantum circuit specializes the QNN into a QCNN (depicted in Fig. \ref{Fig:qcnn}).

\subsection{Quantum encoding}
A transformation, denoted as quantum feature map, performs an encoding $\phi: \mathcal{X} \rightarrow \mathcal{H}$, where $\mathcal{X}$ is a dataset and $\mathcal{H}$ is a Hilbert space. This quantum feature map is implemented as a unitary transformation $U_\phi(x)$ applied to the initial state $|0\rangle^{\otimes n}$  to produce $U_\phi(x)|0\rangle^{\otimes n} = |\phi(x)\rangle$, where $n$ is the number of qubits.

There are various methods to encode the data~\cite{schuld2018supervised}. In this work, two methods are considered to encode the classical data $x_i$ in quantum states.

In the first method, denoted as amplitude encoding, an item $x$ is encoded in a quantum state by associating the normalized features with the probability amplitudes of the quantum state. Given $x = (x_1, x_2, ..., x_N)^T$, with $N = 2^n$, the quantum state is as follows:
\begin{equation}
    U_\phi(x): x \in \mathbb{R}^N \rightarrow |\phi(x)\rangle = \frac{1}{||x||}\sum_{i=1}^{N}{x_i|i\rangle},
\end{equation}
where $|i\rangle$ is the $i$th computational basis state.
The advantage of this method is that it can represent exponentially many classical data using a low number of qubits, but the quantum circuit that implements the encoding has a depth which usually grows as $O(poly(N))$.

The second method, denoted as angle encoding, encodes $N$ features into the rotation angles of $N$ qubits. Given the data $x = (x_1, x_2, ..., x_N)^T$, this method encodes $x$ as
\begin{equation}
     U_\phi(x): x \in \mathbb{R}^N \rightarrow |\phi(x)\rangle = \otimes_{i=1}^{N}{cos(x_i)|0\rangle + sin(x_i)|1\rangle}.
\end{equation}
This encoding uses $O(N)$ qubits.

\subsection{Variational Quantum Circuit}
In a QCNN, the variational quantum circuit is composed of two parts: a convolutional layer and a pooling layer.

In classical CNN, the convolutional layer applies a filter (a small mask) to the images. This filter is slid over different positions of the input image; for each position, an output value is generated by performing the scalar product between the mask and the covered image. Its goal is to accurately identify patterns such as angles in the image. In QCNN, this is done using a circuit applied to pairs of qubits (Fig.~\ref{Fig:qcnn}). Single-qubit rotations are used to rotate the qubit independently and the controlled rotations are used to create the entanglement between the qubits and to correlate their states. The entanglement is useful for identifying complex patterns and it can increase the accuracy of the network.

\begin{figure*}
    \centering
    \includegraphics[width=1.0\textwidth]{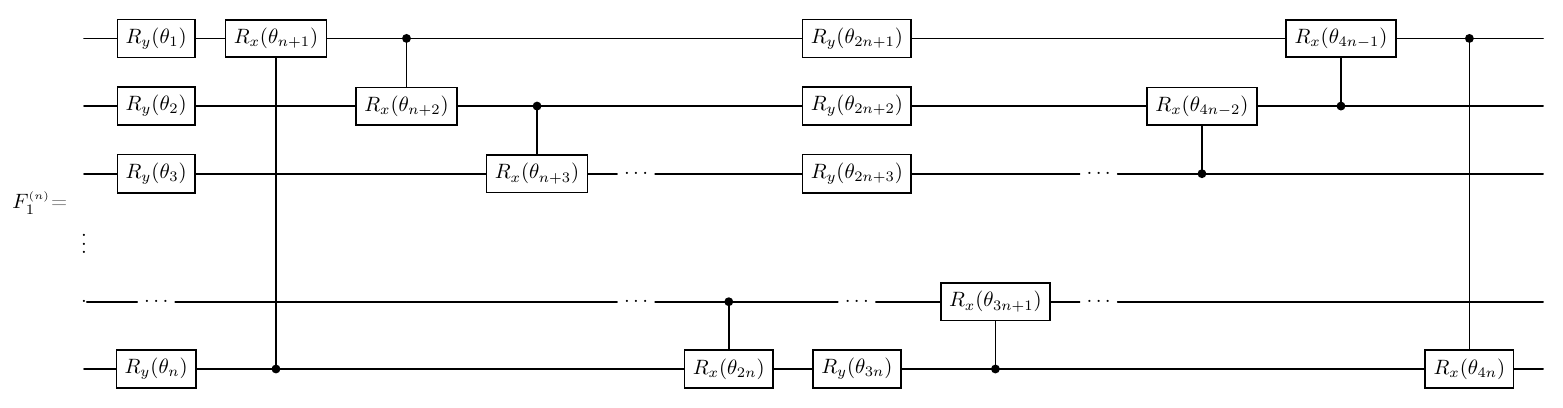}
    \hspace{0.1\textwidth}
    \includegraphics[width=0.7\textwidth]{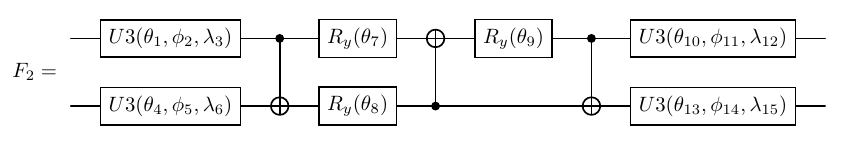}
    \caption{Parameterized quantum circuit used in the convolutional layer. $R_i(\theta)$ is a rotation around the $i$ axis by an angle of $\theta$. $U3(\theta, \phi, \lambda)$ is an arbitrary single-qubit gate that can be expressed as $U3(\theta, \phi, \lambda) = R_z(\phi)R_x(-\pi/2)R_z(\theta)R_x(\pi/2)R_z(\lambda)$.}
    \label{Fig:conv}
\end{figure*}

In this work, the convolutional filter is constructed as follows. First of all, the circuit $F^{(n)}_1$ of Fig.~\ref{Fig:conv}, acting on $n$ qubits, is applied prior to the convolutional filter to perform an initial preprocessing of the images. While in the study~\cite{bokhan2022multiclass}, the preprocessing circuit was applied once before all convolution and pooling layers, here, the circuit is utilized as a preconvolutional circuit, i.e., it is applied before each convolutional layer, to enhance the performance of the quantum neural network. The circuit was proposed in~\cite{schuld2020circuit} by Schuld et al. and it has a low-depth and it achieves good values of expressibility and entangling capability, as shown in~\cite{sim2019expressibility}. For this reason, the circuit can cover the space effectively (expressibility) and can find both short and long-range correlations (entanglement).
After that, the circuit $F_2$ of Fig.~\ref{Fig:conv}, which represents an arbitrary $SU(4)$ gate~\cite{vatan2004optimal, maccormack2022branching}, is applied to pairs of adjacent qubits, following the structure of~\cite{cong2019quantum}.

\begin{figure}
    \centering
    \includegraphics[width=0.4\textwidth]{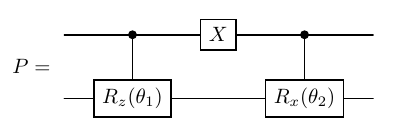}
    \caption{Parameterized quantum circuit used for implementing the pooling operation~\cite{hur2022quantum}.}
    \label{Fig:pooling}
\end{figure}

In classical CNN, each convolutional layer is followed by a pooling layer, which reduces the dimensionality of the data being processed. In QCNN, this is performed by applying a 2-qubit gate and then one qubit is traced out after the controlled gate action. The pooling gate consists of two controlled rotations, following the work of~\cite{hur2022quantum} (Fig.~\ref{Fig:pooling}). It applies two rotations: $R_z(\theta_1)$ and $R_x(\theta_2)$, respectively activated when the control qubit is $1$ and $0$. Then, the control qubit is removed after the operations.

In~\cite{hur2022quantum}, the pooling layer reduced the number of qubits by half using one layer. Conversely, in this work, the pooling operation is applied only twice (instead of, for example, four times in the first layer) per layer, enabling the addition of another convolutional layer before measurement, as depicted in Fig.~\ref{Fig:qcnn}.

The QCNN for 10-class classification is depicted in Fig.~\ref{Fig:qcnn}. To conduct 4/6/8-class classification, an additional pooling layer is added before measurement. This layer aims to reduce the number of qubits to 2 for the 4-class scenario and to 3 for 6 and 8 classes.

\subsection{Optimization}
The parameters of the variational quantum circuit are updated to minimize the cost function. The cross-entropy cost function is used. Given $C$ classes, with indexes from $0$ to $C - 1$, the loss is defined as
\begin{equation}
    \begin{split}
        l(x, y) = L = \{l_1, l_2, ..., l_N\}^T, \\
        l_n = -\sum_{c=0}^{C-1}{ \log\frac{\exp(x_{n,c})}{\sum_{i=0}^{C-1}{\exp(x_{n,i})}}y_{n,c} },
    \end{split}  
\end{equation}
where $x$ is the input, $y$ is the target and $N$ spans the minibatch dimension. 

The training process of the QCNN is carried out using classical method. In particular, the Adam optimizer is used to train the parameters $\theta$.

\section{Results}
\label{sec:results}
\begin{figure*}[ht!]
    \centering
    \begin{subfigure}[b]{0.45\textwidth}
        \centering
        \begin{tikzpicture}
        \begin{axis}[
            title={4-class classification accuracy},
            xlabel={Epochs},
            ylabel={Accuracy},
            xmin=0, xmax=10,
            ymin=0, ymax=100,
            xtick={2, 4, 6, 8, 10},
            ytick={0,20,40,60,80,100},
            legend pos=south east,
            ymajorgrids=true,
            grid style=dashed,
            grid=major,
            width=\textwidth,
            height=5cm,
            xlabel style={yshift=10pt},
        ]
        
        \addplot[
            color=red,
            ]
            coordinates {
            (0,0)(2, 83.5)(4,83.5)(6,84)(8,85)(10,85)
            };

        \addplot[
            color=blue,
            ]
            coordinates {
            (0,0)(2, 84)(4,84.5)(6,85)(8,85)(10,85)
            };
            \legend{Amplitude, Angle}
            
        \end{axis}
        \end{tikzpicture}
        \label{fig:sub4-class}
    \end{subfigure}
    \vspace{-15pt}
    \begin{subfigure}[b]{0.45\textwidth}
        \centering
        \begin{tikzpicture}
        \begin{axis}[
            title={6-class classification accuracy},
            xlabel={Epochs},
            ylabel={Accuracy},
            xmin=0, xmax=10,
            ymin=0, ymax=100,
            xtick={2, 4, 6, 8, 10},
            ytick={0,20,40,60,80,100},
            legend pos=south east,
            ymajorgrids=true,
            grid style=dashed,
            grid=major,
            width=\textwidth,
            height=5cm,
            xlabel style={yshift=10pt},
        ]
        
        \addplot[
            color=red,
            ]
            coordinates {
            (0,0)(2, 66)(4,71)(6,71)(8,72)(10,72)
            };

        \addplot[
            color=blue,
            ]
            coordinates {
            (0,0)(2, 68)(4,68)(6,68)(8,68)(10, 67.5)
            };
            \legend{Amplitude, Angle}
            
        \end{axis}
        \end{tikzpicture}
        \label{fig:sub6-class}
    \end{subfigure}
    \vspace{-15pt}
    \begin{subfigure}[b]{0.45\textwidth}
        \centering
        \begin{tikzpicture}
       \begin{axis}[
            title={8-class classification accuracy},
            xlabel={Epochs},
            ylabel={Accuracy},
            xmin=0, xmax=10,
            ymin=0, ymax=100,
            xtick={2, 4, 6, 8, 10},
            ytick={0,20,40,60,80,100},
            legend pos=south east,
            ymajorgrids=true,
            grid style=dashed,
            grid=major,
            width=\textwidth,
            height=5cm,
            xlabel style={yshift=10pt},
        ]
        
        \addplot[
            color=red,
            ]
            coordinates {
            (0,0)(2, 67.5)(4,69)(6,69.5)(8,70)(10,70)
            };

        \addplot[
            color=blue,
            ]
            coordinates {
            (0,0)(2, 58.5)(4,58)(6,58)(8,58.5)(10,58)
            };
            \legend{Amplitude, Angle}
            
        \end{axis}
        \end{tikzpicture}
        \label{fig:sub8-class}
    \end{subfigure}
    \begin{subfigure}[b]{0.45\textwidth}
        \centering
        \begin{tikzpicture}
        \begin{axis}[
            title={10-class classification accuracy},
            xlabel={Epochs},
            ylabel={Accuracy},
            xmin=0, xmax=10,
            ymin=0, ymax=100,
            xtick={2, 4, 6, 8, 10},
            ytick={0,20,40,60,80,100},
            legend pos=north west,
            ymajorgrids=true,
            grid style=dashed,
            grid=major,
            width=\textwidth,
            height=5cm,
            xlabel style={yshift=10pt},
        ]
        
        \addplot[
            color=red,
            ]
            coordinates {
            (0,0)(2, 56)(4,57)(6,57.5)(8,57)(10,57)
            };

        \addplot[
            color=blue,
            ]
            coordinates {
            (0,0)(2, 44)(4,44)(6,43)(8,43)(10,43.5)
            };
            \legend{Amplitude, Angle}
            
        \end{axis}
        \end{tikzpicture}
        \label{fig:sub10-class}
    \end{subfigure}
    \captionsetup{skip=1pt}
    \caption{Comparison of the classification accuracy between the QCNN using amplitude and angle encoding. The evaluation involves increasing the number of epochs and varying the number of classes, with learning rate equal to 0.01. In the 4-class case, classes 0 to 3 are considered, while for 6 classes, classes from 0 to 5 are taken into account, and for 8 classes, classes 0 to 7 are included.}
    \label{fig:epochs_encoding}
\end{figure*}
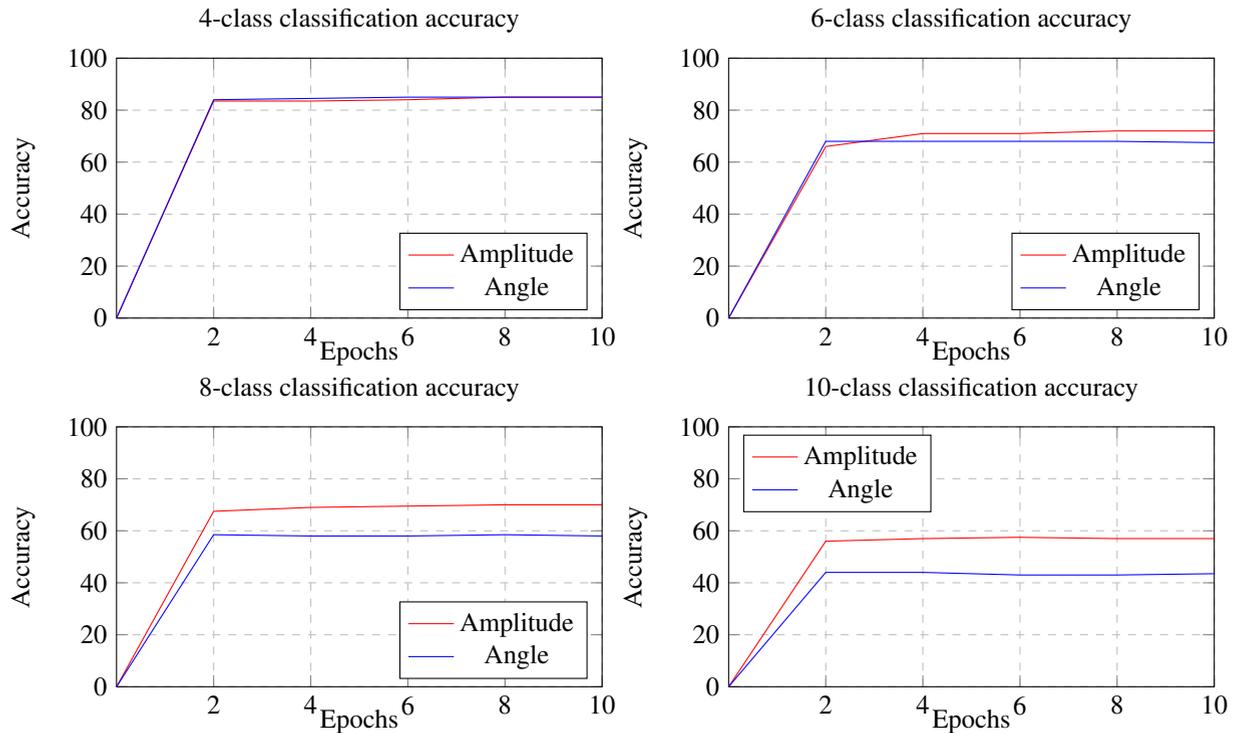

The library PennyLane was used for the implementation of the proposed QCNN. The tests were performed using the simulator provided by the same library.

The optimization of the quantum circuit is based on gradient descent, specifically utilizing Adam, with varying learning rate: 0.01, 0.001 and 0.0005 across 10 epochs. Performance evaluation involves measuring accuracy on the complete test set, while precision, recall and F1 score are calculated to help identify the classes where the network faces recognition challenges.

The implemented QCNN was tested on the MNIST dataset for the classification of handwritten digits. This dataset is composed of 60000 training example images and 10000 testing images of size $28x28$ with 10 classes.
The tests were executed with 4 (0-3), 6 (0-5), 8 (0-7) and 10 classes. 

The images are normalized to one since only that type of vector can be used by the amplitude encoding algorithm. Also, features with values closer to each other should lead to better performance for machine learning algorithms. After the normalization, the images are processed with a dimensionality reduction technique, called Principal Component Analysis (PCA), to reduce the size of the images from $28x28$ to $16x16$, in the case of amplitude encoding. In this way, the images consist of 256 pixels and they can be encoded into a 8-qubit state space. If the angle embedding is used, only the 8 most significant features are retained. 

The 10-class classification involves measuring 4 qubits, where only the states from $|0000\rangle$ to $|1001\rangle$ are taken into consideration and the other states are discarded. For 4/6/8-class classifications, an additional pooling layer is added before measurement. In the 4-class classification, states from $00$ to $11$ are considered. For the 6-class case, states from $000$ to $101$ are included, and for the 8-class scenario, all the states from $000$ to $111$ are taken into account.

\begin{table}[htbp]
  \centering
  \caption{Comparison between angle and amplitude encoding after 10 epochs and with different learning rate.}
    \begin{tabular}{ |p{1cm}|p{1cm}||p{2cm}|p{2cm}|  }
     \hline
     \multicolumn{4}{|c|}{Classification Accuracy} \\
     \hline
     Classes & learning rate & Angle & Amplitude \\
     \hline
     4   & $0.01$ & $86\%$    & $\bm{85\%}$ \\
     6   & $0.01$ & $68\%$  & $\bm{72.2\%}$ \\
     8   & $0.01$ & $58\%$  & $\bm{70\%}$ \\
     10  & $0.01$ & $46\%$& $\bm{57\%}$ \\
     \hline
     4   & $0.001$ & $85\%$    &$80\%$ \\
     6   & $0.001$ & $62.5\%$  & $63.5\%$ \\
     8   & $0.001$ & $56\%$  & $60.5\%$ \\
     10  & $0.001$ & $46\%$& $49\%$ \\
     \hline
     4   & $0.0005$ & $84.7\%$    &$79\%$ \\
     6   & $0.0005$ & $60\%$  & $60\%$ \\
     8   & $0.0005$ & $50.8\%$  & $60\%$ \\
     10  & $0.0005$ & $45\%$& $48\%$ \\
     \hline
    \end{tabular}
  \label{tab:encoding}
\end{table}

The results using angle and amplitude encoding are presentend in Table~\ref{tab:encoding}. In the 4-class classification, both embeddings exhibit very similar performance. However, as the number of classes increases, the performance of angle encoding deteriorates more quickly compared to amplitude encoding. Fig.~\ref{fig:epochs_encoding} shows the accuracy of the model as the number of epochs increases and the number of classes varies. In the plots, it can be observed that accuracy nearly reaches its peak value consistently after two epochs across different executions. The accuracy tends to approach these values within the first two epochs. The accuracy improvement become marginal with additional epochs. 

\begin{table}[htbp]
 \centering
 \caption{Comparison between classical CNN and QCNN with 0.01 of learning rate and amplitude encoding, after 10 epochs.}
    \begin{tabular}{ |p{1cm}||p{2cm}|p{2cm}|  }
     \hline
     \multicolumn{3}{|c|}{Classification Accuracy} \\
     \hline
     Classes & CNN & QCNN \\
     \hline
     4    & $90\%$    &$85\%$ \\
     6 &   $69\%$  & $72\%$ \\
     8 &   $50\%$  & $70\%$ \\
     10  & $38\%$& $57\%$ \\
     \hline
    \end{tabular}
  \label{tab:compare}
\end{table}

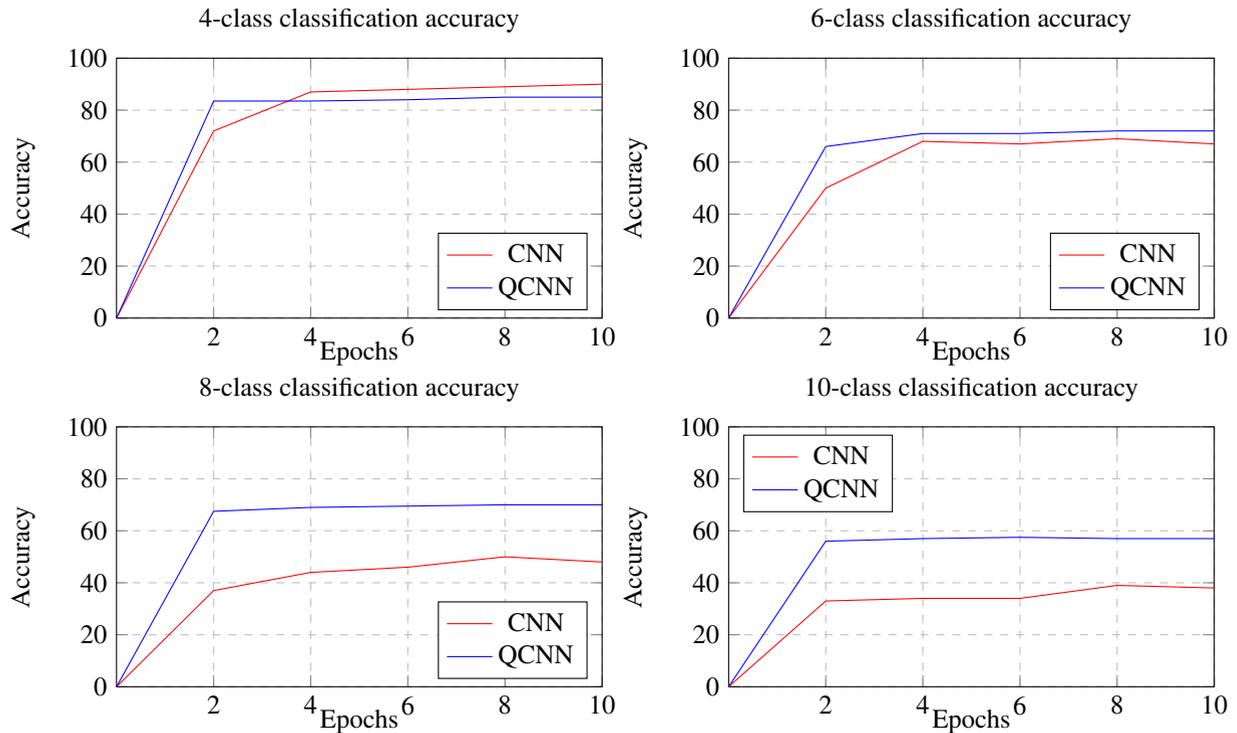
\begin{figure*}[ht!]
    \centering
    \begin{subfigure}[b]{0.45\textwidth}
        \centering
        \begin{tikzpicture}
        \begin{axis}[
            title={4-class classification accuracy},
            xlabel={Epochs},
            ylabel={Accuracy},
            xmin=0, xmax=10,
            ymin=0, ymax=100,
            xtick={2, 4, 6, 8, 10},
            ytick={0,20,40,60,80,100},
            legend pos=south east,
            ymajorgrids=true,
            grid style=dashed,
            grid=major,
            width=\textwidth,
            height=5cm,
            xlabel style={yshift=10pt},
        ]
        
        \addplot[
            color=red,
            ]
            coordinates {
            (0,0)(2,72)(4,87)(6,88)(8,89)(10,90)
            };

        \addplot[
            color=blue,
            ]
            coordinates {
            (0,0)(2, 83.5)(4,83.5)(6,84)(8,85)(10,85)
            };
            \legend{CNN, QCNN}
            
        \end{axis}
        \end{tikzpicture}
        \label{fig:sub4-class2}
    \end{subfigure}
    \vspace{-15pt}
    \begin{subfigure}[b]{0.45\textwidth}
        \centering
        \begin{tikzpicture}
        \begin{axis}[
            title={6-class classification accuracy},
            xlabel={Epochs},
            ylabel={Accuracy},
            xmin=0, xmax=10,
            ymin=0, ymax=100,
            xtick={2, 4, 6, 8, 10},
            ytick={0,20,40,60,80,100},
            legend pos=south east,
            ymajorgrids=true,
            grid style=dashed,
            grid=major,
            width=\textwidth,
            height=5cm,
            xlabel style={yshift=10pt},
        ]
        
        \addplot[
            color=red,
            ]
            coordinates {
            (0,0)(2, 50)(4,68)(6,67)(8,69)(10, 67)
            
            };

        \addplot[
            color=blue,
            ]
            coordinates {
            (0,0)(2, 66)(4,71)(6,71)(8,72)(10,72)
            };
            \legend{CNN, QCNN}
            
        \end{axis}
        \end{tikzpicture}
        \label{fig:sub6-class2}
    \end{subfigure}
    \vspace{-15pt}
    \begin{subfigure}[b]{0.45\textwidth}
        \centering
        \begin{tikzpicture}
       \begin{axis}[
            title={8-class classification accuracy},
            xlabel={Epochs},
            ylabel={Accuracy},
            xmin=0, xmax=10,
            ymin=0, ymax=100,
            xtick={2, 4, 6, 8, 10},
            ytick={0,20,40,60,80,100},
            legend pos=south east,
            ymajorgrids=true,
            grid style=dashed,
            grid=major,
            width=\textwidth,
            height=5cm,
            xlabel style={yshift=10pt},
        ]
        
        \addplot[
            color=red,
            ]
            coordinates {
            (0,0)(2, 37)(4,44)(6,46)(8,50)(10,48)
            };

        \addplot[
            color=blue,
            ]
            coordinates {
            (0,0)(2, 67.5)(4,69)(6,69.5)(8,70)(10,70)
            };
            \legend{CNN, QCNN}
            
        \end{axis}
        \end{tikzpicture}
        \label{fig:sub8-class2}
    \end{subfigure}
    \begin{subfigure}[b]{0.45\textwidth}
        \centering
        \begin{tikzpicture}
        \begin{axis}[
            title={10-class classification accuracy},
            xlabel={Epochs},
            ylabel={Accuracy},
            xmin=0, xmax=10,
            ymin=0, ymax=100,
            xtick={2, 4, 6, 8, 10},
            ytick={0,20,40,60,80,100},
            legend pos=north west,
            ymajorgrids=true,
            grid style=dashed,
            grid=major,
            width=\textwidth,
            height=5cm,
            xlabel style={yshift=10pt},
        ]
        
        \addplot[
            color=red,
            ]
            coordinates {
            (0,0)(2, 33)(4,34)(6,34)(8,39)(10,38)
            };

        \addplot[
            color=blue,
            ]
            coordinates {
            (0,0)(2, 56)(4,57)(6,57.5)(8,57)(10,57)
            };
            \legend{CNN, QCNN}
            
        \end{axis}
        \end{tikzpicture}
        \label{fig:sub10-class2}
    \end{subfigure}
    \captionsetup{skip=1pt}
    \caption{
    Comparison of the classification accuracy of CNN and QCNN. The evaluation involves increasing the number of epochs and varying the number of classes, with learning rate equal to 0.01. In the 4-class case, classes 0 to 3 are considered, while for 6 classes, classes from 0 to 5 are taken into account, and for 8 classes, classes 0 to 7 are included.}
    
    \label{fig:compare_epochs}
\end{figure*}

The comparison between the QCNN and the CNN is presented in Table~\ref{tab:compare}. 
The CNN is constructed using a number of parameters similar to the number of the QCNN. For 10-class classification, the QCNN has $105$ parameters, while an additional $2$ parameters are added when there are 4/6/8 classes, since another pooling layer is needed. 
The CNN is structured as follows: it comprises $3$ convolutional layers, mirroring the setup of the QCNN. Each layer contains only one kernel (or filter). The first convolution has a kernel size of $8x8$, the second $5x5$ and the third $3x3$. It has Tanh as activation function and the pooling operation employed is average pooling, with a kernel size of $2$. Lastly, a fully-connected layer is appended with a number of output features corresponding to the specific number of classes (4, 6, 8 or 10). The CNN's parameter count varies: it has $109$ parameters for 4 classes, $113$ for 6 classes, $117$ for 8 classes, and $121$ for 10 classes.

The QCNN has better performance for 6, 8 and 10 classes. It achieves 72\% of accuracy for 6 classes, 70\% for 8 classes and 57\% for 10 classes, while classical CNN attains 69\%, 50\% and 38\% respectively. Even if the CNN is trained for an additional 90 epochs, it still fails to match the performance of the QCNN.

In Fig.~\ref{fig:compare_epochs}, a detailed depiction of how the performance of CNN and QCNN varies over epochs is provided. The plots demonstrate that while the CNN achieves better results when it sees the same data multiple times, the QCNN does not show significantly better results. By incrementing the number of epochs for the classical CNN, while a better result is obtained with 4 classes, with 6, 8 and 10 classes the CNN cannot reach the same result of the QCNN.

\begin{figure*}[h!]
    \centering
    \begin{subfigure}[b]{0.45\textwidth}
        \centering
        \begin{tikzpicture}
        \begin{axis}[
            title={4-class classification accuracy},
            xlabel={Epochs},
            ylabel={Accuracy},
            xmin=0, xmax=100,
            ymin=0, ymax=100,
            xtick={2, 6, 10, 20, 50, 80, 100},
            ytick={0,20,40,60,80,100},
            legend pos=south east,
            ymajorgrids=true,
            grid style=dashed,
            grid=major,
            width=\textwidth,
            height=5cm,
            xlabel style={yshift=10pt},
        ]
        
        \addplot[
            color=red,
            ]
            coordinates {
            (0,0)(2,24)(6,27)(10,49)(20,50)(50,57)(80,77)(100,82)
            };

        \addplot[
            color=blue,
            ]
            coordinates {
            (0,0)(2, 56)(6,66)(10,67)(20,79)(50,80)(80,83)(100,84)
            };
            \legend{CNN, QCNN}
            
        \end{axis}
        \end{tikzpicture}
        \label{fig:sub4-class3}
    \end{subfigure}
    \vspace{-15pt}
    \begin{subfigure}[b]{0.45\textwidth}
        \centering
        \begin{tikzpicture}
        \begin{axis}[
            title={6-class classification accuracy},
            xlabel={Epochs},
            ylabel={Accuracy},
            xmin=0, xmax=100,
            ymin=0, ymax=100,
            xtick={2, 6, 10, 20, 50, 80, 100},
            ytick={0,20,40,60,80,100},
            legend pos=south east,
            ymajorgrids=true,
            grid style=dashed,
            grid=major,
            width=\textwidth,
            height=5cm,
            xlabel style={yshift=10pt},
        ]
        
        \addplot[
            color=red,
            ]
            coordinates {
            (0,0)(2, 15)(6,33)(10, 38)(20,50)(50,51)(80,59)(100,60)
            
            };

        \addplot[
            color=blue,
            ]
            coordinates {
            (0,0)(2, 39)(6,57)(10,64)(20,70)(50,73)(80,73)(100,71)
            };
            \legend{CNN, QCNN}
            
        \end{axis}
        \end{tikzpicture}
        \label{fig:sub6-class3}
    \end{subfigure}
    \vspace{-15pt}
    \begin{subfigure}[b]{0.45\textwidth}
        \centering
        \begin{tikzpicture}
       \begin{axis}[
            title={8-class classification accuracy},
            xlabel={Epochs},
            ylabel={Accuracy},
            xmin=0, xmax=100,
            ymin=0, ymax=100,
            xtick={2, 6, 10, 20, 50, 80, 100},
            ytick={0,20,40,60,80,100},
            legend pos=south east,
            ymajorgrids=true,
            grid style=dashed,
            grid=major,
            width=\textwidth,
            height=5cm,
            xlabel style={yshift=10pt},
        ]
        
        \addplot[
            color=red,
            ]
            coordinates {
            (0,0)(2, 22)(6,30)(10,33)(20,37)(50,43)(80,38)(100,41)
            };

        \addplot[
            color=blue,
            ]
            coordinates {
            (0,0)(2, 42)(6,47)(10,54)(20,63)(50,65)(80,65)(100,68)
            };
            \legend{CNN, QCNN}
            
        \end{axis}
        \end{tikzpicture}
        \label{fig:sub8-class3}
    \end{subfigure}
    \begin{subfigure}[b]{0.45\textwidth}
        \centering
        \begin{tikzpicture}
        \begin{axis}[
            title={10-class classification accuracy},
            xlabel={Epochs},
            ylabel={Accuracy},
            xmin=0, xmax=100,
            ymin=0, ymax=100,
            xtick={2, 6, 10, 20, 50, 80, 100},
            ytick={0,20,40,60,80,100},
            legend pos=north west,
            ymajorgrids=true,
            grid style=dashed,
            grid=major,
            width=\textwidth,
            height=5cm,
            xlabel style={yshift=10pt},
        ]
        
        \addplot[
            color=red,
            ]
            coordinates {
            (0,0)(2, 10)(6,13)(10,19)(20,22)(50,32)(80,40)(100,34)
            };

        \addplot[
            color=blue,
            ]
            coordinates {
            (0,0)(2, 27)(6,36)(10,40)(20,44)(50,48)(80,50)(100,52)
            };
            \legend{CNN, QCNN}
            
        \end{axis}
        \end{tikzpicture}
        \label{fig:sub10-class3}
    \end{subfigure}
    \captionsetup{skip=1pt}
    \caption{
    Comparison of the classification accuracy of CNN and QCNN with only 250 samples per class. The evaluation involves increasing the number of epochs and varying the number of classes, with learning rate equal to 0.01. In the 4-class case, classes 0 to 3 are considered, while for 6 classes, classes from 0 to 5 are taken into account, and for 8 classes, classes 0 to 7 are included.}
    
    \label{fig:250samplesPlots}
\end{figure*}
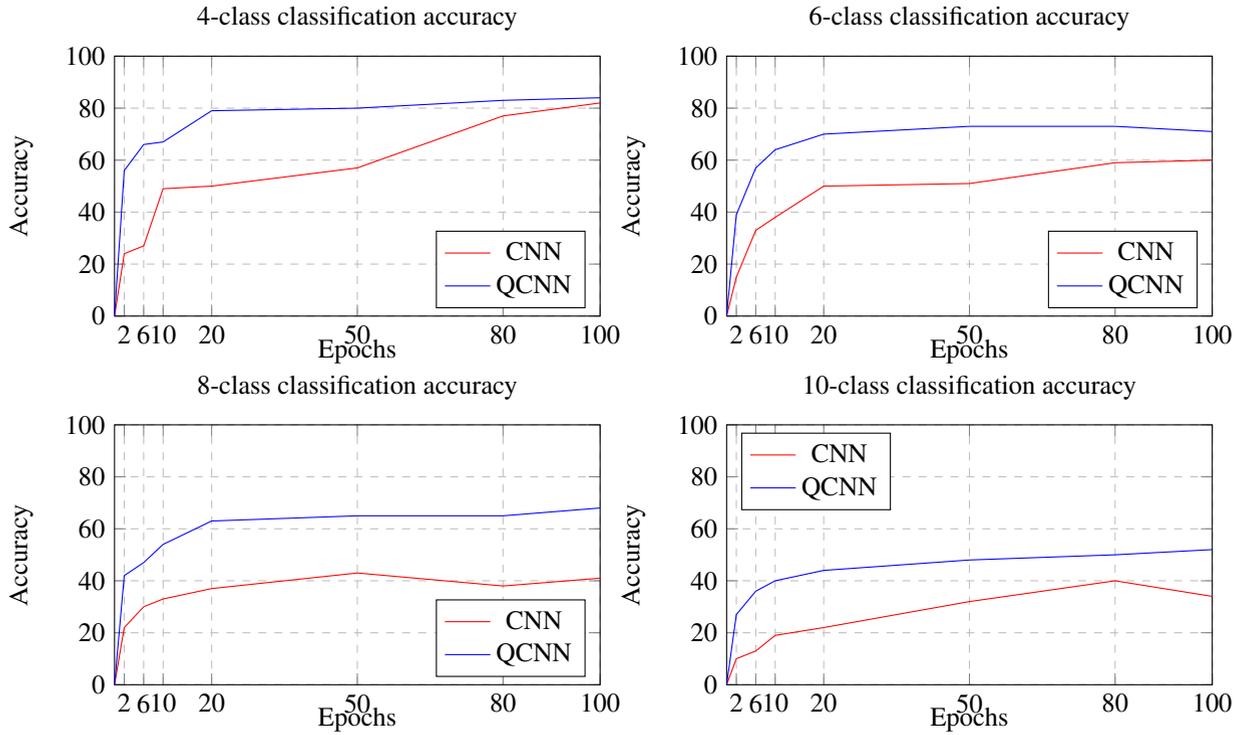

As a final experiment, the QCNN was tested with only 250 samples per class and the results were compared with those of the CNN. In Fig.~\ref{fig:250samplesPlots}, the accuracy over 100 epochs is depicted for both neural networks. The plots show that the QCNN almost achieves the same level of performance as when using the entire dataset, whereas the classical CNN fails to achieve comparable results with a limited number of samples. 

These results can help in the future to study how generalization and overfitting work in the quantum scenario. In classical scenarios, networks typically achieve better performance when provided with more samples. Conversely, in the quantum case, the quantum networks demonstrate the capability to achieve comparable result with a smaller number of samples.
Another interesting aspect is that the QCNN achieves or nearly achieves peak performance with the entire training set within a limited number of epochs, showing minimal improvements with more epochs. This contrasts with classical networks, which heavily rely on an increased number of epochs for notable enhancements.

In Table~\ref{tab:10C}, the results achieved by the QCNN for each class are presented. There are classes where the network achieves good results (classes: 0, 1, 2, 3 and 7). However, there are classes, such as class 5, 6 and 9, where the network struggles to achieve accurate recognition.

\begin{table}[htbp]
  \centering
  \caption{Precision, recall and F1 score of the QCNN for 10 classes, after 10 epochs. Amplitude encoding is used and the learning rate is equal to 0.01.}
    \begin{tabular}{ |p{1cm}||p{2cm}|p{2cm}|p{2cm}|p{2cm}|  }
     \hline
     Class & Precision & Recall & F1 score\\
     \hline
     0    & $61\%$    &$67\%$ &   $64\%$\\
     1 &   $65\%$  & $93\%$   &$76.5\%$ \\
     2  & $60\%$& $66\%$&  $63\%$ \\
     3     &$51\%$ & $68\%$&  $58\%$ \\
     4     &$67\%$ & $37\%$&  $48\%$ \\
     5     &$27\%$ & $11\%$&  $15.5\%$ \\
     6     &$44\%$ & $54\%$&  $54\%$ \\
     7     &$55\%$ & $80.5\%$&  $65\%$ \\
     8     &$59\%$ & $34\%$&  $43\%$ \\
     9     &$72\%$ & $18\%$&  $29\%$ \\
     \hline
    \end{tabular}
  \label{tab:10C}
\end{table}

All data are available from the corresponding authors upon reasonable request.

\section{Conclusion}
\label{sec:conclusion}
Classification is particularly relevant to IR, as it is used in various subtasks of the search pipeline.
With the potential of quantum computers to handle efficiently an ever-increasing amount of data, new perspectives on classification algorithms are opening up.

In this paper, a QCNN was presented for solving multi-class classification problems on classical data.
The QCNN has better performance compared to the classical counterpart for 6, 8 and 10 classes. Furthermore, the proposed quantum neural network nearly achieves the results of~\cite{bokhan2022multiclass} with a significantly lower number of parameters. 

The tested QCNN uses a parameters count equal to $105$ in the 10-classes scenario, while with 4, 6 and 8 classes, an additional $2$ parameters are added since another pooling layer is appended before measurement. The QCNN is characterized by the following classification accuracy: $86\%$ in the 4-class scenario, $72\%$ with 6 classes, $70\%$ with 8 classes and $57\%$ with 10 classes.

In the future, the proposed QCNN framework may be enhanced by modifying the convolutional and the pooling layers. The pooling layer could be modified to transmit information about the state of the qubit traced out to multiple qubits, extending beyond the single adjacent qubit.

Further potential improvements involve the measurement process. In the scenario with 10 classes, the network performs measurements on 4 qubits, leaving 6 states unused for predicting the 10 classes. This can introduce noise in the classification process, potentially impacting performance. A similar situation arises with 6 classes. One solution could be to utilize 10 qubits during measurement, associating each qubit with one of the 10 classes. Each qubit would be set to state 1 when an image of the respective class is inputted.

Understanding generalization and overfitting is an important aspect. Research on this topic has already initiated, as seen in ~\cite{gil2023understanding, caro2022generalization,peters2023generalization}. This paper can help to understand how generalization and overfitting work in the quantum scenario, because it seems like the quantum networks can achieve optimal results in generalization with a fewer number of samples.

Another possible future development is to add more kernels (filters) to observe their impact on network performance, since in the classical scenario this often leads to enhanced performance.

\section*{Acknowledgement}
Michele Amoretti acknowledges financial support from the European Union – NextGenerationEU, PNRR MUR project PE0000023-NQSTI.
This research benefits from the HPC (High Performance Computing) facility of the University of Parma, Italy.

\bibliographystyle{IEEEtran}
\bibliography{references}

\end{document}